\documentclass[%
 reprint,
superscriptaddress,
msmath,amssymb,
amsfonts,amsmath,floatfix
aps,
prb,
]{revtex4-1}
\pdfoutput=1
\usepackage{graphicx}
\usepackage{amsmath}
\usepackage{amssymb}
\usepackage{dsfont}
\usepackage{tensor}
\usepackage{color}
\usepackage{soul}
\usepackage{ulem}
\usepackage{cancel}
\usepackage{upgreek}
\usepackage{braket}
\usepackage{hyperref}
\usepackage{blindtext}
\usepackage{dsfont}
\usepackage{amsmath}
\usepackage{amssymb}
\usepackage{ulem}
\usepackage[letterpaper,left=1.5cm,right=1.5cm,top=1.6 cm,bottom=2.5cm]{geometry}
\setstcolor{red}

\begin{document}
\pdfoutput=1
\preprint{APS/123-QED}


\title{Understanding Contagion Dynamics through Microscopic Processes in Active Brownian Particles}

\author{Ariel Norambuena}
\affiliation{Centro de Investigaci\'on DAiTA Lab, Facultad de Estudios Interdisciplinarios, Universidad Mayor, Chile}
\author{Felipe Valencia}
\affiliation{Centro de Investigaci\'on DAiTA Lab, Facultad de Estudios Interdisciplinarios, Universidad Mayor, Chile}
\affiliation{Centro para el Desarrollo de la Nanociencia y la Nanotecnología, CEDENNA, Avda. Ecuador 3493, Santiago 9170124, Chile}
\author{Francisca Guzm{\'a}n-Lastra}
\email{francisca.guzman@umayor.cl}
\affiliation{Centro de Investigaci\'on DAiTA Lab, Facultad de Estudios Interdisciplinarios, Universidad Mayor, Chile}
\affiliation{Escuela de Data Science, Facultad de Estudios Interdisciplinarios, Universidad Mayor, Chile}

\date{\today}

\begin{abstract}
Together with the universally recognized SIR model, several approaches have been employed to understand the contagious dynamics of interacting particles. Here, Active Brownian particles (ABP) are introduced to model the contagion dynamics of living agents that spread an infectious disease in space and time. Simulations were performed for several population densities and contagious rates. Our results show that ABP not only reproduces the time dependence observed in traditional SIR models, but also allows us to explore the critical densities, contagious radius, and random recovery times that facilitate the virus spread. Furthermore, we derive a first-principles analytical expression for the contagion rate in terms of microscopic parameters, without the assumption of free parameters as the classical SIR-based models. This approach offers a novel alternative to incorporate microscopic processes into the analysis of SIR-based models with applications in a wide range of biological systems.
\end{abstract}

\maketitle

\section*{Introduction}

Mathematical models and computational calculations provide powerful scientific tools to understand and predict future scenarios associated with viral propagation dynamics. Nowadays, the global impact of COVID-19 demands new paradigms to explore novel theoretical models used in disciplines like physics, chemistry, biology, ecology, mathematics, and computational science, to improve our understanding of pandemic events. Even more, motivated by the hope of enriching our knowledge of the complex contagion dynamics in living agents. Historically, infectious diseases have been modeled using the SIR model~\cite{McKendrick1927} (and its variations) using coupled non-linear differential equations, which include phenomenological rates to describe the contagion, recuperation, death, quarantine or lock-down. Nevertheless, a more realistic model must consider the mobility of infectious particles and particle density within its environment. In this direction, self-propelled particles~\cite{Peruani}, the random motion of non-interacting particles~\cite{Rodriguez2019}, cellular automaton~\cite{Bernhard2008,Neri2020}, dynamical density functional theory approach~\cite{Wittkowski2020}, and reaction-diffusion models~\cite{Murray2003,Sokolov2006} have been proposed to introduce the spatial motion of infectious particles. As a matter of universality, random diffusion models are intuitive and are extensively used to describe a wide range of biological processes ranging from bacteria motion to animal movement. Thus, as active matter lies at the core of almost all biological processes, it emerges as an excellent and non explored candidate to describe the contagion dynamics of moving particles. \par

Active matter (AM) affects the organization and collective behavior of different living organisms on all length scales, ranging from cytoskeleton on the nanoscale through cheeps on the mesoscale~\cite{sumpter2006principles,shaebani2020computational,klotsa2019above,ramaswamy2010mechanics,vicsek2012collective}. Since the work of Viscek et al.~\cite{vicsek1995novel}, for self-driven particles, modeling collective behaviors have been possible following a series of rules for particle interactions, such as alignment, polarization, repulsion, group sensing, among others~\cite{chate2008modeling,marchetti2013hydrodynamics, bechinger2016active,berdahl2018collective}. These interactions often give rise to the understanding of unexpected phenomena such as turbulence, giant fluctuations, rectification, and self-organization~\cite{linkmann2019phase,dabelow2019irreversibility,chate2006simple, nishiguchi2017long,dunkel2013fluid,lozano2016phototaxis,tierno2008magnetically} and at the same time they reproduce what we observe in nature. Living organisms move on fluids media, and their dynamics can be characterized by the Reynolds number $Re=v_0L/\mu$, where $v_0$ is particle's velocity, $L$ is the body length and $\mu$ the dynamic viscosity of the fluid. This dimensionless number compares inertial forces with viscous forces giving rise to different limits where either the modeling and particle behaviors seem to be listed. At low Reynolds number,  $Re\ll 1$, viscous forces dominate over inertial ones, which is often observed in the nano and the micro scales. In this regime, there has been a theoretical, numerical and experimental effort to model, control and understand active matter for the promising applications in medicine, mining industry, intelligent crops, and ecology~\cite{figueroa2020coli,mathijssen2018nutrient,ramos2020bacteria,pietzonka2019autonomous,mallory2018active,costerton1987bacterial,denissenko2012human,dwyer2012bioflotation}. \par

At low Reynolds number, there has been a theoretical, numerical and experimental effort to model, control and understand active matter for the promising applications in medicine, mining industry, intelligent crops, and ecology~\cite{figueroa2020coli,mathijssen2018nutrient,ramos2020bacteria,pietzonka2019autonomous,mallory2018active,costerton1987bacterial,denissenko2012human,dwyer2012bioflotation}. In this regimen we can model agents as active Brownian particles (ABP). Brownian particles can take up energy from the environment to store it in an internal depot and convert internal energy into kinetic energy~\cite{ebeling1999active} and motion, therefore thermal fluctuations in these systems are dominant~\cite{volpe2014simulation, zottl2016emergent,romanczuk2012active}. ABP has been tested reproducing either biological processes or artificial ones in several studies where it seems that activity and short-range interactions are enough to understand particle-particle and particle-surface interactions \cite{liebchen2019interactions,giomi2013swarming,sanchez2012spontaneous,deblais2018boundaries,schweitzer2000modelling,ebeling1999active}. Nevertheless, in the presence of an external flow or for flagellated microorganisms, more sophisticated models are required to reproduce their behavior ~\cite{li2009accumulation,martinez2018advances,martinez2018emergent,debnath2018hydrodynamic,debnath2019active}. \par

\begin{figure}[ht]
\centering
\includegraphics[width=1 \linewidth]{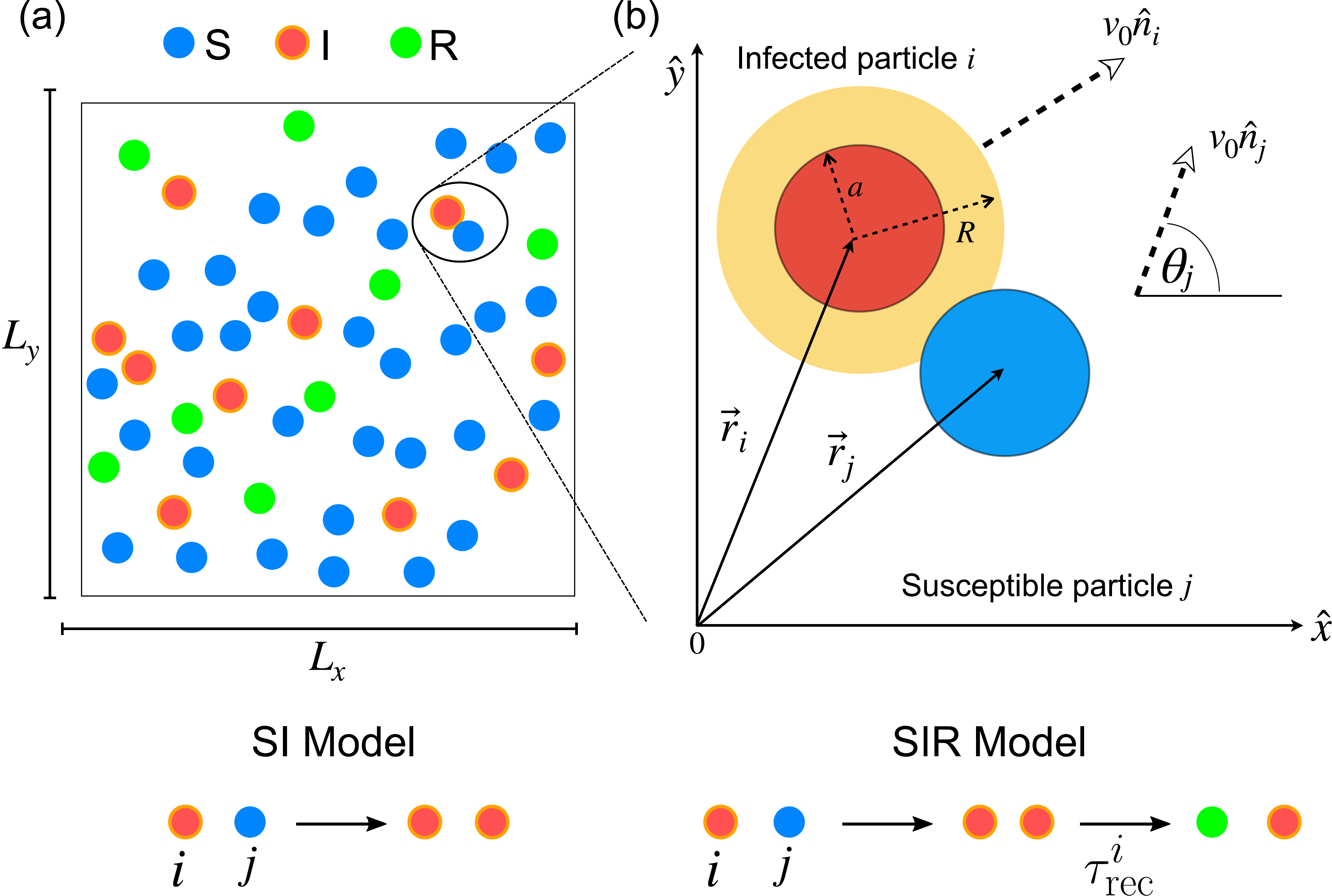}
\caption{Schematic representation of the AM model based on ABP. (a) Sketch of the simulation box: a squared box of size $L_x\times L_y$ with periodic boundary conditions. Initially, we randomly set the initial positions $\vec r_i$ and orientations $\hat n_i$ for all particles $i$. While particles interact more susceptible particles $S$ (in blue) get infected $I$ (in red) yet after a time $\tau_{rec}^{i}$ they get recovered from the infection $R$ (in green). (b) Particle infection: Pair interactions between particle $i$ and $j$. Infected particle $i$ is moving with velocity $v_0\hat n_i$ and given position $\vec r_i$ and interact through the contagion radius $R$ with particle $j$ which is moving with velocity $v_0\hat n_j$ and position $\vec r_j$. }
\label{fig:Figure1}
\end{figure}

AM on the mesoscale has been less explored \cite{klotsa2019above, mathijssen2019collective}. In this regime, inertia and viscous forces are balanced. 
Although living systems in this length scale are plenty such as marine and aerial group of animals, their modeling is less unified since their dynamics depends on the fluid media where they move, and also because now particle interactions get more specific  \cite{mathijssen2019collective, gilpin2017vortex} in function on the target problem.\cite{mathijssen2019collective,gilpin2017vortex}. While the Viseck model or based-agent models are still used to model population dynamics under dry conditions or when the fluid media is air \cite{sepulveda2017wetting,wysocki2020capillary,jafari2019biologically,berdahl2018collective}. \par

Here, we explore infection propagation through active vectors, such as Salmon hatcheries, mosquitos, and mesoscale organisms \cite{martinez2020trapping,ramirez2020multi}. Specifically, in dry active systems, in 2D, we introduce a new AM based model to simulate virus spreading, which lies in the intersection between physics, biology, and computation. In our approach, we introduce $N$ interacting particles following the Langevin equations of a random diffusion process. Moreover, by performing the ensemble average to our AM model, we obtain a similar SIR dynamics, and we derive an alternative microscopic expression for the contagion rate. Our findings show a good agreement between simulation and theory.

\section*{Active Brownian particles}
Let us consider a two-dimensional system composed of $N$ ABP moving at constant speed $v_0$ performing a persistent movement with rotational diffusion $D_R$ in a rectangular box with periodic boundary conditions. 

\begin{figure}[ht]
\centering
\includegraphics[width=1 \linewidth]{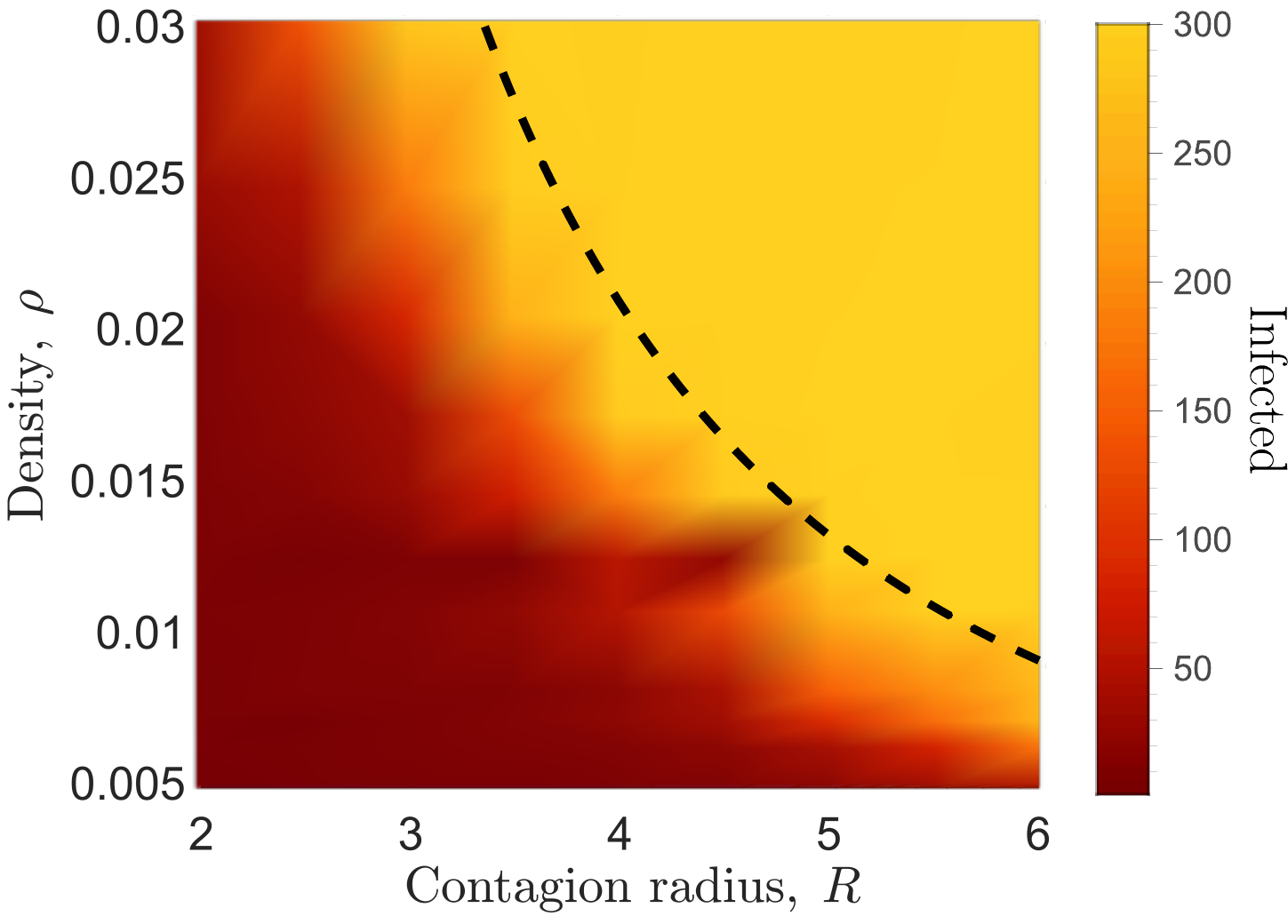}
\caption{Phase diagram for the SI model showing the number of infected particles as a function of the contagion radius $R$ and the particle density $\rho = N/(L_x L_y)$. The dashed black line represent the critical density $\rho_{\rm c} = 1/(\pi R^2)$. For the simulation we consider $N= 300$, $v_0 = 1$, $L_x = L_y$, and $I(0)=1$.}
\label{fig:Figure2}
\end{figure}

Particles are represented by interacting disks of radius $a$ with instantaneous position $\vec{r}_i = x_i\mathbf{e}_{x} + y_i\mathbf{e}_{y}$ and orientation $\theta_i$ respect to the laboratory $x-$axis, where $\mathbf{e}_{x}= (1,0)$, $\mathbf{e}_{y}= (0,1)$ are the unit vectors. For very close encounters particles interact with each other via a Weeks-Chandler-Andersen (WCA) potential to account exclude volume interactions and particle contagion,

\begin{equation}\label{eq.1}
  U_{ij} =  \left\{\begin{array}{cc}
     \displaystyle{4 \varepsilon\left[\left(\frac{r_0}{r_{i j}}\right)^{12}-\left(\frac{r_0}{r_{ij}}\right)^{6}\right]}    &   r_{i j} \leq r_{m}\\
     0    & \mbox{otherwise} 
    \end{array} \right.
\end{equation}

Here, $\varepsilon$ is the interaction potential constant, $r_m$ locates the potential minimum, which is equal to the limit distance between particles $r_0=2a$. Particles diffuses under the combined action of self-propulsion with director vector $\hat n_i=(\cos\theta_i,\sin\theta_i)$ and repulsive forces derived from the pair-repulsive interactions~\eqref{eq.1} avoiding clashes between particles and exploring a confined space. Therefore, we assumed damped particle dynamics, neglecting inertia on particle dynamics, and considering the following set of Langevin equations,

\begin{equation}
\begin{aligned}
&\dot{x}_{i}=-\sum_{j \neq i} F_{i j}^{x}+v_{0} \cos \theta_{i}\\
&\dot{y}_{i}=-\sum_{j \neq i} F_{i j}^{y}+v_{0} \sin \theta_{i}\\
&\dot{\theta}_{i}=\xi_{i}^{\theta},\label{eqtheta}
\end{aligned}
\end{equation}

where $F_{i j}^{\alpha} = -\left[\nabla U_{ij}\right] \cdot \mathbf{e}_{\alpha}$ are the cartesian components of the force with $\alpha = x,y$. Due to the particles rotational diffusion, the angles $\theta_i$ change randomly according to the Wiener process of~\eqref{eqtheta}, where $\left\langle\xi_{i}^{\theta}(t)\right\rangle=0$ and $\left\langle\xi_{i}^{\theta}(t) \xi_{i}^{\theta}(0)\right\rangle=2 D_{R} \delta(t)$. For an active particle rotational diffusion $D_R$ is related with the medium viscosity and temperature, here we assume it as a constant parameter that takes account particle's exploration of the medium. 

\section*{Active Brownian particles and SI model}

First, we consider a simple SI model where infected $I(t)$ and susceptible $S(t)$ satisfy $I(t) + S(t) = N$. A contagious event occurs when a susceptible particle $i$ is in contact with an infected particle $j$ at a distant $d_{ij} = |\vec{r}_i-\vec{r}_j| \leq R$, where $R$ is the contagion radius, as shown in Fig.~\ref{fig:Figure1}. Also, we assume that infected particles cannot be recovered, we set $I(0) = 1$, and initially all particles are randomly distributed over the area $A = L_x L_y$. For a set of parameters $(N,L_x,L_y,R,v_0)$ we run $N_{\rm sim}$ simulations to compute ensemble average: $I(t) = \sum_{i=1}^{N_{\rm sim}} I_i(t)/N$ and $S(t)  = \sum_{i=1}^{N_{\rm sim}} S_i(t)/N$. \par

As a first computational experiment we simulate the response of the system by changing the particle density $\rho = N/(L_x L_y)$ and the contagion radius, by setting $N=300$, $N_{\rm sim}=100$, $v_0 = 1$ and $L_x = L_y$. In Fig.~\ref{fig:Figure2}, a color map show the number of infected particles as a function of the contagion rate and the particle density. As expected, in the region of high density and large contagion radius, the infected group saturates reaching its maximum value, \textit{i.e.} $I \approx 300$. More importantly, we observe the existence of a critical density $\rho_{\rm c} = 1/(\pi R^2)$ (black dashed line) above which the particles are immediately infected. This motivates a more profound analysis of the microscopic processes related to the contagion dynamics. Using a mean-free-path analysis (see Methods~\ref{MFP} for further details), we obtain the following analytical expression for the contagion rate:

\begin{equation}\label{rate}
    r = \frac{\sqrt{8} \rho R v_0 }{1-\rho/\rho_{\rm crit}},  \quad \quad 0 \leq \rho \leq \rho_{\rm crit}.
\end{equation}

In the low-density regime, $\rho \ll \rho_{\rm crit}$, we obtain a linear scaling $r \approx \sqrt{8} \rho R v_0$. Also, our model predicts a singularity at $\rho = \rho_{\rm crit}$ for which $r \rightarrow \infty$. In such a case, $r$ diverges, revealing that all particles are instantaneously infected. One critical observation is the dimensional-dependent nature of the contagion rate in our model. For instance, if we $N$ particles moving in a volume $V$, the mean-free-path analysis predicts a 3D contagion rate $r^{3\rm D} = \pi \rho^{3\rm D}R^2\langle v_{\rm rel}\rangle/(1-\rho^{3\rm D}/\rho^{3\rm D}_{\rm crit})$, where $\rho^{3\rm D} = N/V$, $\rho^{3\rm D}_{\rm crit} = 1/(4/3\pi R^3)$, and $\langle v_{\rm rel}\rangle$ is the average relative velocity between particles. Therefore, our active matter model predicts that distancing between infected particles is more critical in a three-dimensional system since $r^{3\rm D} \propto R^2$. The latter can be crucial in biological systems where a 3D movement is present during the contagion dynamics~\cite{deblais2018boundaries, giomi2013swarming, sumpter2006principles, jeanson2005self, garnier2009self}. \par

\begin{figure}[ht]
\centering
\includegraphics[width=1 \linewidth]{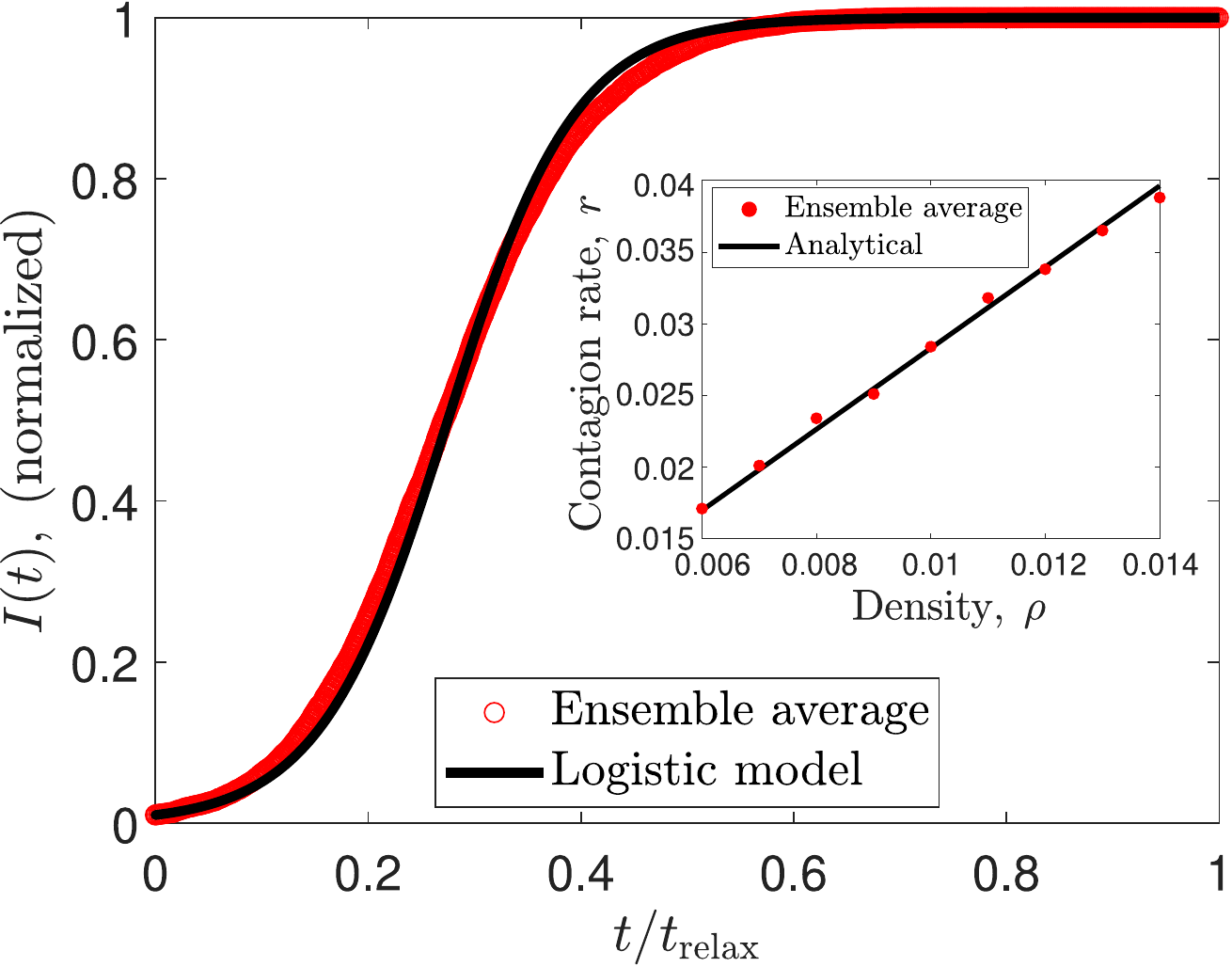}
\caption{Time evolution of the infected group for the SI model. The red circles are numerical simulations of the Langevin equations after calculating the ensemble average. The solid black line is the solution of~\eqref{Lotka-Volterra}. For the simulation we use $N = 100$, $R=1$, $L_x = L_y = 100$, and $v_0 =1$. Here, $t_{\rm relax}$ is the relaxation time required to find the stationary state of the system. The inset plot show the contagion rate as a function of the particle density, where we compare the analytical expression derived in~\eqref{rate} (solid line) with our simulation (red circles). For the simulation we use $N=100$, $N_{\rm sim} = 100$ $R=1$, $L_x = L_y$, and $v_0 =1$. }
\label{fig:Figure3}
\end{figure}

In minimal models for active matter, such as the ABP model with exclude volume interactions, we have fundamental mechanisms to observe and understand the emergence of complex dynamics such as the clustering formation or bimodal phase separation while varying the particles activity or density in these systems \cite{redner2013structure, stenhammar2014phase, fily2014freezing, cates2015motility}. This two-phase separation between a solid-like phase and a gas-like phase has also been observed in experiments with carbon-coated Janus particles which are self-propelled artificial microswimmers~\cite{buttinoni2013dynamical, palacci2013living} and surprisingly in social behavior such as circle pits in heavy metal concerts where in very dense systems more active particles forms clusters that move around the space where less active particles are jammed \cite{silverberg2013collective}. This type of phase separation, in the scenario of the virus propagation, could be relevant since cluster formation might be treated as density gradients in the space induced by particles attracted to hot spots in dilute or dense systems. Then, this two-phase system can be used to study the space and time dynamics of particles forced to quarantine in groups or on their city hall while some rangers continue moving in the space between clusters. In this case, we expect that the contagion rate $r,\; r^{3\rm D}$, which is density-dependent, would be measured and accordingly used for novel mechanisms of infection that until now are not described by standard epidemic models \cite{paoluzzi2020information}. \par

\begin{figure}[ht]
\centering
\includegraphics[width=1 \linewidth]{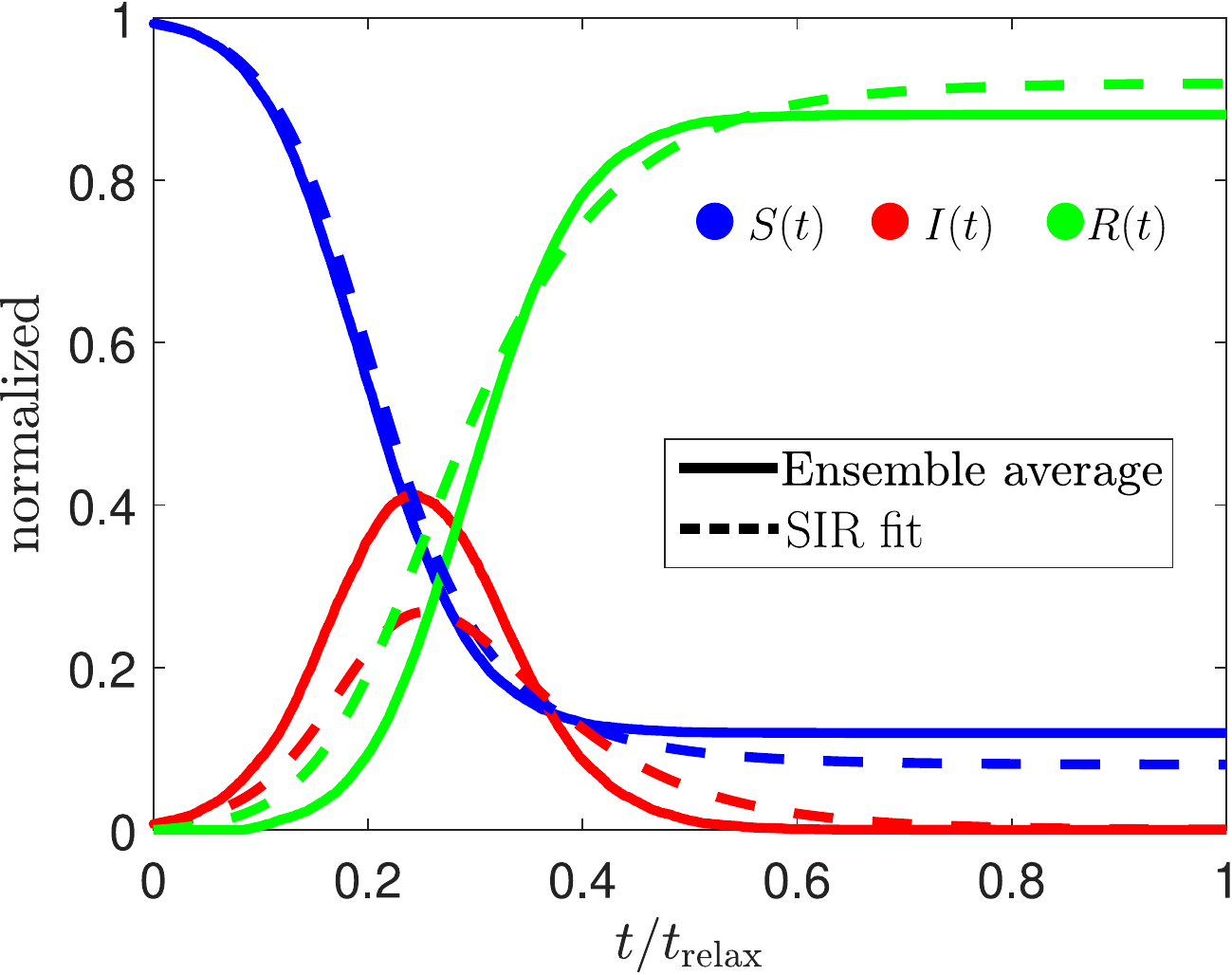}
\caption{Comparison between our SIR model and the best fit obtained by optimizing the parameters $\alpha$ and $\beta$. For the simulation we consider one initial infected particle, $I(0) =1$ and a random recovery time $\tau_{\rm rec}^{i} \in [30,50]$. For the numerical calculations we use $N=150$, $R=1$,$L_x = L_y = 100$, and $v_0 = 1$.}
\label{fig:Figure4}
\end{figure}

Now, we shall establish the connection between our microscopic contagion rate given in~\eqref{rate} and the characteristic epidemic curve for the SI model. At each discrete time $t_n = n\Delta t$ ($n \in \mathds{N}$ and $\Delta t >0$), the number of infected varies according to the Markovian model $I_{n+1} = I_n + p_n S_n$, where $p_n = (r \Delta t)(I_n/N)$ and $S_n = N - I_n$ are the contagion probability and number of susceptible at time $t_n$, respectively. As a consequence, in the continuum limit, the curve $I(t)$ evolves according to ($\Delta t \rightarrow 0$):

\begin{equation}\label{Lotka-Volterra}
\dot{I} = rI\left(1-\frac{I}{N} \right),  \quad S(t) = N-I(t).
\end{equation}

The above equations can be written as $\dot{S} = -r I S /N$ and $\dot{I}= r I S /N$, which is the standard SI model. The logistic function $I(t) = I(0)Ne^{r t}/[(N-I(0))+I(0)e^{rt}]$ gives the analytic solution of~\eqref{Lotka-Volterra}. To support our previous observations, in Fig.~\ref{fig:Figure3}, we plot a comparison between the infected curve $I(t)$ obtained from the ensemble average procedure and the logistic model given above. Here, we consider a system with $N=100$ particles in a square box with lengths $L_x = L_y =100$, contagion radius $R=1$, and velocity $v_0 = 1$. We observe a good agreement between the theory and simulations, revealing that one initial contagion grows logistically if the recovered group is neglected. However, a small asymmetry of the analytical logistic model is observed in Fig.~\ref{fig:Figure2}. One suggestive approach is to fit the ensemble average with the generalized logistic model or Richard's model given by $\dot{I} = r I^p[1-(I/N)^q]$ ($0 \leq p \leq 1$) which has been used in COVID-19 pandemic curves~\cite{Vasconcelos}. This could be useful for biological systems showing logistic-like behaviors with more involved microscopic dynamics.  \par

Furthermore, in the inset of Fig.~\ref{fig:Figure3}, we compare the microscopic expression for the contagion rate defined in~\eqref{rate} and the predicted rate obtained in our simulations. We recover the predicted linear dependence of the contagion rate in terms of the particle density, which validates our microscopic model. More generally, the contagion rate given in~\eqref{rate} can be also estimated for a system with different velocities by using $ r = 2 \rho R \langle v_{\rm rel} \rangle /(1-\rho/\rho_{\rm crit})$, where $\langle v_{\rm rel} \rangle$ is the average relative velocity between particles.

\section*{Active Brownian particles and SIR model}

Now, we include the recovered group $R(t)$ into the dynamics. In such a case, the total number of particles satisfy $S(t) + I(t) + R(t) = N$. First, we assume that the recovered group cannot be infected again, that is, particles gain immunity after the contagion process. Second, we neglect deaths since we are interested in the propagation itself. Third, we introduce a random recovery time $\tau_{\rm rec}^{i}$ for each particle ($i=1,...,N$) such that $\tau_{\rm rec}^{i} \in [\tau_{\rm min}, \tau_{\rm max}]$. Here, $\tau_{\rm min}$ and $\tau_{\rm max}$ are the minimum and maximum recovery in our simulations, respectively. \par 

\begin{figure}[ht]
\centering
\includegraphics[width=1 \linewidth]{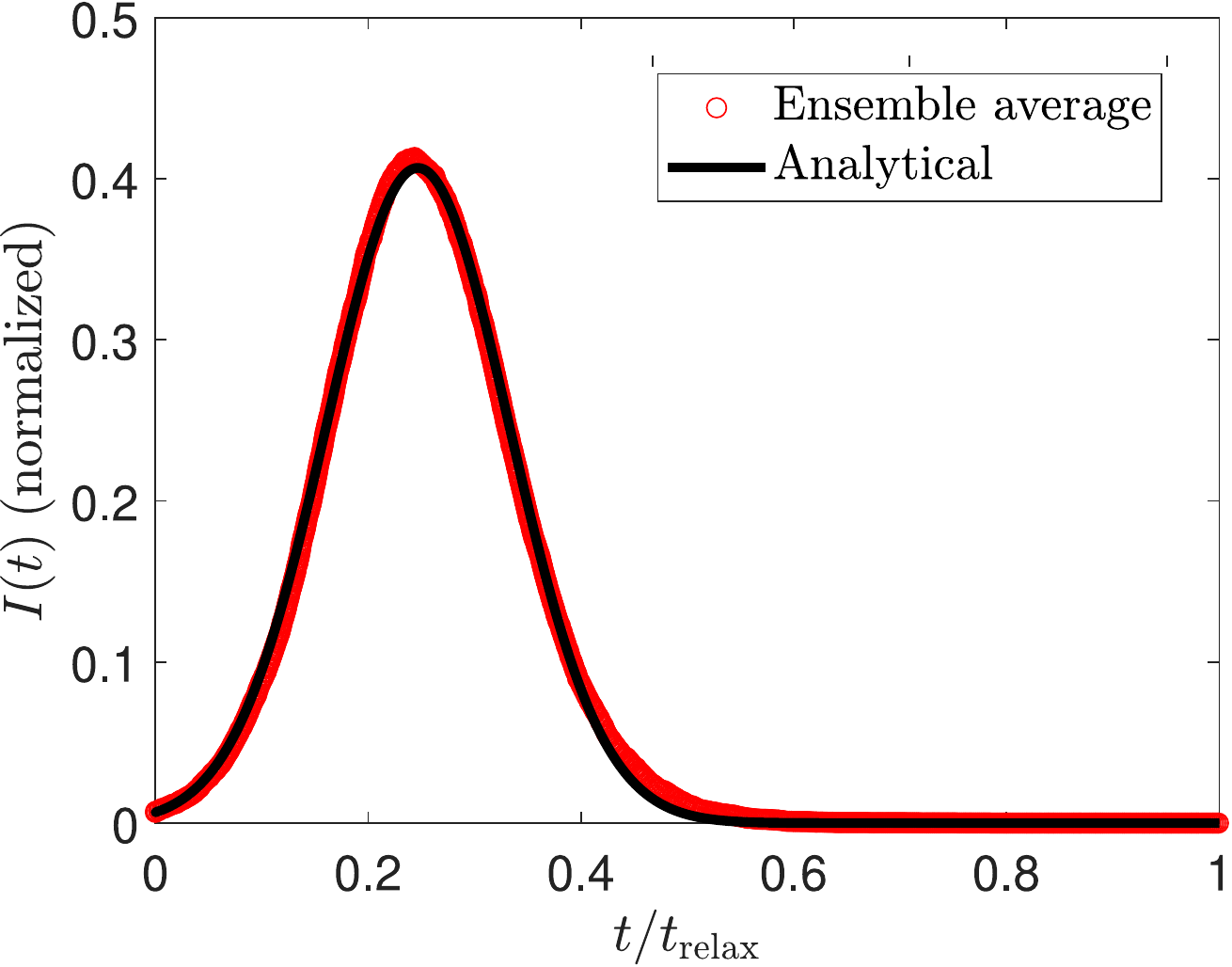}
\caption{Infected curve and analytical Gaussian prediction for the SIR model. For the numerical calculations we use $N=150$, $R=1$,$L_x = L_y = 100$, $v_0 = 1$, $I(0) =1$, and $\tau_{\rm rec}^{i} \in [30,50]$}
\label{fig:Figure5}
\end{figure}

We compare our simulations with the conventional SIR model, which is described by the set of differential equations $\dot{S} = -\alpha  IS$, $\dot{I} = \alpha  I S - \beta I$, $\dot{R} = \beta  I$, where $\alpha$ and $\beta$ are the infection and recovery rates, respectively~\cite{McKendrick1927}. We can find the optimal parameters $\alpha$ and $ \beta$ that improves the fit between the SIR model and our simulations. In Fig.~\ref{fig:Figure5}, we observe a comparison between our simulations (ensemble average) and the SIR fit (dashed lines). In general, we numerically corroborate that our model cannot be fully explained in terms of the standard SIR model. In particular, the SIR model predicts an asymmetry curve for $I(t)$ and the stationary states differs with our calculations. Our simulations shows a symmetric curve for the infected group, which has been previously observed in Ref.~\cite{Tsironis}. However, using our microscopic point of view we can use the relations $\alpha = r/N$, where $r$ is given in~\eqref{rate} and $\beta = 1/T_{\rm prom}$ with $T_{\rm prom} = (\tau_{\rm min}+\tau_{\rm max})/2$ being the average recovery time. Moreover, the differential equation $\dot{I} = r  IS /N -  I/T_{\rm prom}$ can be solved by noting that the relevant contribution to the product $IS$ comes from the region where $S(t)$ has a linear dependence. Note that we have a microscopic basic reproduction number defined as $R_0 = r T_{\rm prom}/N = \sqrt{8} R v_0 T_{\rm prom}/[A(1-\rho/\rho_{\rm crit})]$ for which $\left. \dot{I}\right|_{t=0} > 0$ if $S(0) >R_0$. By using the approximation $S(t) = S_0 - m t$ into the dynamics of $I(t)$, we found the following Gaussian curve:

\begin{equation}\label{Gaussian-I}
    I(t) = I(0)e^{\left({t_0 \over \sqrt{2}\sigma}\right)^2}e^{-\left({t-t_0 \over \sqrt{2}\sigma}\right)^2},
\end{equation}

where $t_0 = (r S_0  - N/T_{\rm prom})/(r m)$ is the position of the peak and $\sigma =  [N/(rm)]^{1/2}$ is the width of the Gaussian function in terms of microscopic parameters. In Fig.~\ref{fig:Figure5}, we observe the good agreement between our simulations and the Gaussian model given in~\eqref{Gaussian-I}. On the one side, the maximum number of infected is estimated as $I_{\rm max} \approx I(0)\mbox{exp}[(t_0/(\sqrt{2}\sigma))^2]$, and thus the ratio $t_0/\sigma$ is critical. In the low-density regime, we obtain $I_{\rm max} \propto \mbox{exp}[R v_0/A]$ illustrating that the contagion radius, available area, and velocity of particles strongly impact the maximum number of infected during the dynamics. On the other hand, the scaling $\sigma \propto [R v_0/A]^{-1/2}$, tell us that any reduction of the maximum number of infected implies a flattened effect on the curve $I(t)$, as expected in the standard SIR model.

Further improvements or extensions of the current model can be performed by considering the incubation time, different particle velocities, time-dependent densities to model lock-down, or by including particle interactions modeled with microscopic pedestrian models~\cite{Teknomo}.

\section*{Conclusions}

Active Matter simulations were conducted to study the virus propagation phenomena. Our results show that active Brownian particles can successfully reproduce the universally accepted SIR contagious curves. Additionally, by controlling contagious radii and particle density, we can observe the optimal conditions favoring the spread of viruses. Theoretically, the SIR model assumes several empirical parameters in order to describe the contagious dynamics. Here, we introduce first-principles analytical expression that successfully reproduces the results observed in the active matter simulations in terms of controllable microscopic parameters. Besides, our expression qualitatively recovers the SIR based models, but present a better agreement with the numerical simulations. \par

Here active matter simulations have been employed to study the temporal and spatial contagious dynamics. Although our study focuses mainly on particle density and contagious rate, several parameters as recovering time, particle velocity,  boundary conditions, obstacles, among others, deserve to be studied. We expect that active matter simulations could be a useful tool to study optimal conditions for infection propagation on several systems such as Salmon hatcheries, mosquitos, or human contagious in close ambients, like shopping centers, hospitals, industries among many others.

\section*{Methods}
\subsection{Brownian Dynamics Simulations in the overdamped limit}
We performed Brownian dynamics simulations for $N=300$ disk particles bounded in a squared box of area $A=L_x\times L_y$ with periodic boundary conditions. Particles are settled initially at random positions and orientation following a uniform distribution. Particles move according to Langevin equations~\eqref{eqtheta} with a rotational diffusion given by $D_R=1$ [rad$^2$/s], where we set a new position and orientation for each particle using the Euler iteration method with a time step $dt=10^{-3}$. Since the particle dynamics is non-deterministic and particle encounters determine the contagious rate, we performed $100$ different numerical simulations starting with a different random configuration. Particles perform pair-hard core interactions via the WCA which sets particle size $a=1$ and diameter $2a$. Although this interaction avoids particles overlapping its principal consequence, the particle trajectory deviations imitate living organisms' encounters. Particles also transmit the infection through an instantaneous pair-interaction, which sets a new length parameter on the problem, the contagious radii $R$. Then if the distance between a susceptible particle and an infected particle is less than $R$, the susceptible particle is labeled as infected. We vary the contagious radii from $R=a,\ldots,6$, in steps of $\Delta R=0.5$, and the box length $L=100,\ldots,300$ in increments of $\Delta L=10$~\cite{vrugt2020effects, feng2020influence, guzman2020bioaerosol}.

\subsection{Microscopic contagion rate} \label{MFP}

The microscopic contagion rate can be derived using the concept of mean free path $\lambda$, extensively used in the kinetic theory of gases and also used in Ref.~\cite{Gonzalez}. In this context, $\lambda$ represent the mean distance traveled by ABP between successive encounters with other particle at a distant $d_{ij} = R$. In an active media with $N$ moving particles $\lambda = \sqrt{\langle |\vec{v}_{\rm rel}|^2\rangle}  \tau_c$, with $\vec{v}_{\rm rel}$ and $\tau_c$ being the relative velocity between particles and the mean contagion time, respectively. Here, $\langle ...\rangle$ denote the particle average. Thus, we estimate the contagion rate trough the relation $r = \tau_c^{-1}$. Encounters between ABP's depends on the relative velocity $\vec{v}_{\rm rel} = \vec{v}_i - \vec{v}_j$ ($i\neq j$), from which it follow that $\langle |\vec{v}_{\rm rel}^{\; ij}|^{2}\rangle = \langle v_i^2\rangle + \langle v_i^2\rangle - 2 \langle \vec{v}_i \cdot \vec{v}_j \rangle$. First, we assume uncorrelated particle's velocities yielding to $\langle \vec{v}_i \cdot \vec{v}_j \rangle = 0$. Second, if the WCA potential does not drastically change the speed $v_0$, we approximately obtain that $\langle |\vec{v}_{\rm rel}^{\; ij}|^{2}\rangle \approx 2 v_0^2$ since $\langle v_i^2\rangle \approx v_0^2$. By considering the total area swept for $N$ particles in a time interval $\tau_c$ as $A_{\rm sw} = N(2R\lambda + \pi R^2)$, we define the maximum contagion probability $p_{\rm c} = A_{\rm sw}/A = 1$, and using the relation $\lambda = \sqrt{2}v_0 \tau_c$, we recover the analytical expression of the contagion rate given in~\eqref{rate}.

\begin{acknowledgements}
F. G. L, F. V, and A. N. acknowledges the fruitful discussions with Fernando Crespo. FGL acknowledges Millennium Nucleus Physics of Active Mater of ANID (Chile). FV was supported by the Fondo Nacional  de  Investigaciones  Cient\'ificas  y  Tecnol\'ogicas(FONDECYT, Chile) \#1190662, and \#11190484, and Financiamiento Basal para Centros Cient\'ificos y Tecnol\'ogicos de Excelencia FB-0807, AFB180001. Powered@NLHPC: This research was partially supported by the supercomputing infrastructure of the NLHPC (ECM-02).
\end{acknowledgements}

\section*{Contributions}
F. G. L. and A. N. conceived the research. F. G. L., A. N., and F. V. performed the simulations, A. N. and F. V. analyzed the data. A. N. created the theoretical model. All authors prepared the manuscript, proofread the paper, made comments, and approved the manuscript.

\section*{Interests}
We declare that we have no competing interests.

\section*{Bibliography}
\bibliographystyle{unsrt}
\bibliography{References}

\end{document}